\documentstyle[aps]{revtex}
\begin{document}
\draft
\tighten
\title{Frozen  motion model and dilepton production from 200A GeV
S+W collisions at CERN SPS}
\author{\bf A. K. Chaudhuri\cite{byline}}
\address{ Variable Energy Cyclotron Centre\\
1/AF,Bidhan Nagar, Calcutta - 700 064\\}
\date{\today}
\maketitle

\begin{abstract}
We  have analysed the invariant mass distribution of dileptons in
S+W  collisions  at  200A  GeV  as  measured  by   the   HELIOS-3
collaboration  at CERN SPS. Three scenarios were considered where
the collision lead to formation of (i)  the  quark-gluon  plasma,
(ii)  the  ideal  hot hadron gas and (iii) the viscous hot hadron
gas.  The  space-time  evolution  was  governed  by  the  minimal
extension  of  the Bjorken hydrodynamics. All the three scenarios
indicate excess dileptons in the experiment.
\end{abstract}

PACS number(s): 25.75.Gz, 12.38.Mh\\
\section{Introduction}

In  relativistic  heavy  ion  collisions, the deconfined state of
quarks and gluons (QGP) is expected to be produced.  Lattice  QCD
calculations  also  predict  the formation of QGP. However, until
now there is no conclusive proof that QGP is formed in heavy  ion
collisions.  It  is then essential to consider the possibility of
formation of hot hadronic gas also. The major aim of the  present
day  relativistic  heavy ion experiments is to identify which one
of these two possible states is produced as  the  initial  state.
Dileptons  due  their weak final state interaction are considered
to be ideal probe for investigating  the  initial  state  of  the
fireball.   Several  groups  have  measured  the  invariant  mass
distributions  of  dileptons  at  CERN.  Two   muon   experiments
HELIOS-3/NA34  \cite{ma95}  and  NA38  \cite{ab94}  have measured
dimuons in central S+W collisions in the forward rapidity region.
They observed unaccounted excess of dileptons.  Excess  dileptons
were     also     reported    by    the    CERES    collaboration
\cite{ag95,ts95,wu95}, who have  measured  low  invariant  mass
dileptons  in  the central rapidity region, in central collisions
of  S+Au.  Shuryak  and  Xiong  \cite{sh94}  have  analysed   the
invariant  mass distribution of dileptons as obtained by the NA34
and NA38 collaboration. They assumed that after the  collision  a
mixed  phase  is  produced. They found that in the invariant mass
region 1-2 GeV, the model do not produce sufficient lepton  pairs
to  explain  the  experiments.  They argued that the conventional
model of expansion can  not  explain  the  excess  dileptons  and
proposed  a  slow  expansion model where the mixed phase fireball
expands slowly, during the time  period  of  30-40  fm.  Recently
Srivastava  et  al  \cite{sr96} analysed the CERES as well as the
HELIOS-3 dilepton production data. The CERES data corresponds  to
central  rapidity  range  and  was  analysed  using  the  Bjorken
hydrodynamics. Bjorken hydrodynamics could not be  used  for  the
HELIOS-3  data  as  it lies in the forwards rapidity region. They
followed the treatment of Sinyukov et al \cite{si91} and retained
the distinction between the fluid and particle  rapidities,  with
fluid  rapidities limited between [-1.7,1.7]. They considered two
scenarios, one with the phase transition to the  QGP,  the  other
without  the  phase  transition. It was found that the CERES data
which covered invariant mass range 0.2-1.5 GeV is well  explained
in  both  the  scenario's.  However,  the  HELIOS-3 data could be
explained only in the scenario with  the  phase  transition.  The
scenario  without  the  phase  transition  overpredict  the large
invariant mass data by a factor of $\sim 40$.

In the present paper, we have analysed the HELIOS-3 dilepton mass
distribution  in the frozen motion model \cite{sh94}. The purpose
was to see whether in a different model same conclusion (i.e. the
data  favor  QGP)  can  be  obtained.  The  frozen  motion  model
\cite{sh94} is the minimal extension of Bjorken hydrodynamics and
do  not  demand  plateau  in  the  rapidity  distribution.  It is
appropriate for  the  HELIOS-3  data  which  covers  the  forward
rapidity  region,  where the rapidity distribution do not exhibit
plateau. As in ref. \cite{sr96} two possible scenario's which can
arise after the collision have been considered namely  the  phase
transition  (PT)  scenario  where  QGP  is  formed in the initial
state, and the no phase transition (NPT)  scenario,  where  ideal
hot  hadronic  gas  is  formed. In addition, we have considered a
scenario where viscous hadronic gas  is  formed  in  the  initial
state.  Recently  we  have  shown  that  formation of hot viscous
hadronic gas can explain the WA80 \cite{sa93} preliminary  single
photon   data  for  200  AGeV  S+Au  collisions  reasonably  well
\cite{ch95}. It also describe the $p_T$ distribution  of  neutral
pions  as  obtained  by the WA80 collaboration \cite{ch96}. We do
not reanalyse the CERES data as we are interested to see  whether
the  dilepton  data  are sensitive enough to discriminate between
the different scenario's considered. The analysis  of  Srivastava
et  al \cite{sr96} revealed that the CERES data are not sensitive
to the precise nature of initial state  due  to  low  invariant
mass  coverage.  Also,  in  the  central  rapidity region, frozen
motion model is equivalent to the Bjorken  model.  As  the  CERES
data  corresponds to central rapidity region, any difference from
the analysis of Srivastava et al \cite{sr96} is not expected.

In  the PT scenario it will be assumed that after the collision a
baryon free QGP is formed at the initial  time  $\tau_i$  and  at
temperature  $T_i$.  It expands longitudinally and cools till the
critical temperature $T_c$ is reached and the fluid  enters  into
the  mixed phase. The matter remains in the mixed phase until all
the QGP matter is converted  adiabatically  in  to  the  hadronic
matter.   The   hadronic  matter  then  cools  further  till  the
freeze-out temperature $T_f$ is reached at time $\tau_f$. In  the
NPT scenario's, it will be assumed that after the collision again
a  baryon  free  hot hadronic gas {\em ideal} or {\em viscous} is
formed at initial time $\tau_i$ and at temperature $T_i$. It also
expands longitudinally and cools till the freeze-out  temperature
$T_f$  is reached. As in ref.\cite{sr96}, the hot hadronic matter
will    be    considered    to    be    composed     of     $\pi,
\rho,\omega,\eta$,$\eta$',$\phi$,K  and  $K^*$  as  well as axial
vector mesons $A_1$. As will be shown below, hadron gas with such
a limited number of resonances will require a very large  initial
temperature  to reproduce the experimental multiplicity and it is
difficult to believe in their existence. As in ref.\cite{sr96} we
have neglected transverse expansion. At  SPS  energy,  transverse
expansions are not much \cite{ja83}. Two pion correlation studies
also indicate that if at all there is transverse expansion, it is
small  \cite{al95}.  Srivastava and Sinha \cite{sr94} also showed
that at SPS energy, direct photon production including transverse
expansion differ marginally from the yield obtained assuming only
longitudinal expansion. We thus feel it is justified  to  neglect
transverse expansion.

The  paper is organised as follows: in section 2, we describe the
frozen motion model. Dilepton yields are obtained in  section  3.
The summary and conclusions are given in section 4.

\section{Frozen motion model}

In  the  Bjorken  hydrodynamics  \cite{bj83}, the fluid undergoes
boost-invariant longitudinal expansion. While boost-invariance is
a good approximation in central  rapidity  region,  it  certainly
fails at forward rapidity region. Frozen motion model \cite{sh94}
is  the  minimal  extension  of  the Bjorken hydrodynamics. It is
assumed that at each rapidity, one can consider  a  small  region
where  the  fluid  temperature,  energy  density etc. do not vary
appreciably, and Bjorken hydrodynamics can  be  applied  locally.
This  is  the so called local fluid approximation \cite{vo93} and
reminiscent of local density approximation in  Nuclear  structure
calculations  \cite{ne70}.  This approximation will hold true for
dileptons with  large  transverse  mass  \cite{sa95}.  Space-time
integrated  dilepton yield from a longitudinally expanding system
can be written as \cite{ka86},

\begin{equation}
\frac{dN}{dM^2    dy   d^2p_T}   \approx   \int   d^4x   exp[-M_T
cosh(\Theta-y)/T]
\end{equation}

In the above, $M_T$ is the transverse mass of the lepton pairs, T
is  the temperature and $\Theta$ is the fluid rapidity. For large
values of the transverse mass compared to the temperature, we can
write \cite{mc85},

\begin{equation}
exp[-M_T        cosh(\Theta-y)/T]       \approx       exp[-M_T/T]
exp[-M_T(\Theta-y)^2/2T]
\end{equation}

Thus,  it is seen that if the fluid rapidity density distribution
is a  Gaussian  $\approx  exp(-\Theta^2/2\sigma^2)$  and  $\sigma
>>(T/M_T)^{1/2}$, variation of temperature etc. over the relevant
range  of the fluid rapidity can be ignored \cite{sa95}. In terms
of invariant  mass  (M)  of  the  lepton  pairs,  more  stringent
condition  for  the  validity of local fluid approximation can be
written as,

\begin{equation}
M>>T/\sigma^2
\end{equation}

As will be shown below, at SPS energy for S+W collisions, $\sigma
\sim 1.51$. In the QGP scenario, fluid temperature $\sim$ 200 MeV
is  expected.  The  local  fluid  approximation is then valid for
invariant mass $M>>$0.08 GeV. In the hadronic gas  scenario  much
higher  temperature  (300-400 MeV) is expected. The approximation
will be good only for invariant mass larger than 0.13-0.18 GeV.

In  the  frozen  motion  model \cite{sh94}, it is further assumed
that the fluid rapidity is frozen i.e.  rapidity  of  any  matter
element  remain  unchanged  during  the evolution. The freeze-out
time of the fluid element then can be obtained from  experimental
pion  rapidity  distribution.  Consider  a  fluid  element  in  a
rapidity interval $\Delta \eta_i$ around the  rapidity  $\eta_i$.
For  dileptons  with  large  transverse mass, as indicated above,
energy density $\varepsilon(\tau,\eta_i)$ can be {\em assumed} to
vary negligibly with the rapidity in  that  interval  such  that:
$\varepsilon(\tau,\eta_i)\approx \varepsilon_{\eta_i}(\tau)$. The
space-time  evolution  of the ideal/viscous fluid in the rapidity
interval will be governed by  the  energy  momentum  conservation
equation,  which  for  one  dimensional  similarity  flow  can be
written as \cite{bj83},

\begin{equation}
\frac{d\varepsilon_{\eta_i}}{d\tau}    =    -(\varepsilon_{\eta_i}
+p_{\eta_i}-\frac{4\lambda_{\eta_i}}                      {3\tau}
-\frac{\zeta_{y_i}}{\tau})/\tau \label{1}
\end{equation}

\noindent  where  $\varepsilon_{\eta_i}$ and $p_{\eta_i}$ are the
energy density and pressure of the fluid in the rapidity interval
$\Delta \eta_i$ around the rapidity $\eta_i$.  $\lambda_{\eta_i}$
and  $\zeta_{\eta_i}$  are  the  shear  viscosity  and  the  bulk
viscosity  coefficients  respectively.  To  make   the   notation
simpler,   in  the  following  the  subscript  $\eta_i$  will  be
suppressed with the understanding that, unless otherwise  stated,
the  kinematical  variables  corresponds  to  a rapidity interval
$\Delta \eta_i$ around the rapidity  $\eta_i$,  In  the  scenario
where  the  QGP  or  the  hot  ideal  hadron gas is formed as the
initial state $\lambda=\zeta=0$,  while  in  the  scenario  where
viscous  hadron  gas  is  formed, the viscosity coefficients will
have     definite     values     which     we      take      form
ref.\cite{ch95,da85,ho85}.

\begin{mathletters}
\begin{equation}
\lambda \simeq T/\sigma_{\eta} \label{2}
\end{equation}
\begin{equation}
\zeta=\frac {2}{3} \eta \label{3}
\end{equation}
\end{mathletters}

\noindent  where,  $\sigma_{\eta}$ is the transport cross section
for which a value of 10 mb is used \cite{ch95}.

Eq.\ref{1}  can  be  solved  at  each  fluid  rapidity  given the
equation of state and one boundary condition. We  have  used  the
standard  bag model equation of state $p_q=g_q\pi^2/90T^4-B$ with
$g_q$=47.5 for the QGP phase. For the hadronic  phase,  the  pion
gas  equation  of  state  $p_h=g_h\pi^2/90T^4$ with $g_h$=6.8 was
used \cite{sr96}. The mixed phase was described  by  the  Maxwell
construct  $p_q(T_c)=p_h(T_c)$, which also gives the bag constant
B.

In  the  local  fluid  approximation,  initial temperature of the
fluid depends on the fluid rapidity $T_i=T_i(\eta)$  \cite{vo93}.
For  the  ideal  QGP/hadronic  fluids  for  a  given initial time
$\tau_i$, $T_i(\eta)$ can be obtained  by  equating  the  entropy
density  with the experimental pion multiplicity at that rapidity
\cite{hw85}.

\begin{equation}
T^3_i(\eta)\tau_i=\frac{1}{\pi R^2_A}        \frac{c}{4a_{q,h}}
{(\frac{dN}{dy})}_{y=\eta_i} (b=0) \label{4}
\end{equation}

\noindent  where  $c=2\pi^4/45\zeta(3)$,$a_{q,h}=g_{q,h}\pi^2/90$
and  b=0  corresponds  to  central  collisions.  $R_A$   is   the
transverse  radius  of  the  system.  Eq.\ref{4}  thus  maps  the
experimental   rapidity   distribution   into    a    temperature
distribution.  For  viscous hadronic fluid, eq.\ref{4} can not be
used to obtain $T_i(\eta)$ as entropy is generated in  the  flow.
However,  the  pion  multiplicity  can  be equated with the final
entropy density to obtain the  freeze-out  time  $(\tau_f(\eta)$)
for a given freeze-out temperature $(T_f)$ \cite{ch95},

\begin{equation}
T^3_f\tau_f(\eta)=\frac{1}{\pi R^2_A}        \frac{c}{4a_h}
{(\frac{dN}{dy})}_{y=\eta_i} (b=0) \label{5}
\end{equation}

\noindent   and   the  evolution  equation  can  be  solved  with
($T_f,\tau_f(\eta)$)   to   obtain   the   initial    temperature
$T_i(\eta)$  .  We  note that eq.\ref{5} is obtained if the fluid
flow is convoluted with the thermal distribution and is valid for
both the ideal and the viscous flow.

Now  the  information  on  $dN/d\eta$  vs $\eta$ for the HELIOS-3
experiment is not available. We then proceed with the  assumption
that  the  $dN/d\eta$ vs $\eta$ for S+W and that for S+Au are the
same. In fig. 1, we have shown  the  pseudorapidity  distribution
for  the  charged particles as obtained by the WA80 collaboration
for central S+Au collisions \cite{wa80}. Also shown is a  fit  to
the data by a Gaussian,

\begin{equation}
dN_{ch}/d\eta = \alpha exp(\eta-{\bar \eta})^2/2\sigma^2)
\end{equation}

\noindent   with,   $\alpha=161.48$,   ${\bar   \eta}$=2.63   and
$\sigma$=1.51.

Neglecting   the   difference   between   the  rapidity  and  the
pseudorapidity distributions  (for  large  $y$  they  are  nearly
same), the rapidity distribution $dN/d\eta$ for S+W collisions in
the  HELIOS-3 experiments is obtained by multiplying the above by
1.5 ( we are assuming that for every  charged  pair  there  is  a
neutral  one).  We  note that we may be overestimating $dN/d\eta$
for S+W as we are neglecting the mass difference between Au and W
ions. Thus we may be overestimating the freeze-out time and  also
the  initial  temperature  of  the  fluid.  In fig.2, the initial
temperature  of  the  fluid,  at  the  (canonical)  initial  time
$\tau_i$=1  fm,  in the rapidity range 3.7-5.2, which corresponds
to the  HELIOS-3  experiment  is  shown.  The  critical  and  the
freeze-out  temperatures  were  taken as 160 MeV and 140 MeV. For
the fluid elements with $y> 4.2$, dN/dy is such that, in  the  PT
scenario, the fluid can not exist in a {\em pure} QGP phase. {\em
They  were  assumed  to  be  formed  in  the  mixed  phase,  with
appropriate QGP and hadron gas fraction}.  Thus  in  the  present
discussion,  in  the  phase  transition  scenario,  the  fluid is
essentially  formed  in  a  mixed  phase  \cite{sh94}.  In   this
scenario,  variation  of  $T_i$  over the rapidity range is quite
small (176 -160 MeV).  Much  larger  $T_i$  is  obtained  if  the
initial  state  is the hadron gas. Thus if {\em ideal} hadron gas
is formed $T_i$ varies between 337 MeV  (at  $\eta$=3.7)  to  226
MeV(at  $\eta$=5.2). Initial temperature is reduced if the hadron
gas is {\em viscous}. $T_i$ then varies between 291-161 MeV. This
is because entropy is generated in a viscous flow. In  fig.3,  we
have  shown  the  corresponding  hadronic  density. For the ideal
hadron gas, initial density varies between 4.2-1.2 hadrons/$fm^3$
in the rapidity range considered. It is difficult to  believe  in
the  existence  of  hadronic  gas  with  density  as large as 4.2
hadron/$fm^3$. For example the close pack density of hadronic gas
composed {\em solely} of pions is  $\sim$2.6  pions/$fm^3$.  With
resonances,  it  will  still  be  lower.  It is thus difficult to
believe    that    hadronic    gas    comprising    of     $\pi$,
$\rho$,$\omega$,$\eta$,$\eta\prime$,$\phi$,  $K$, $K^*$ and $A_1$
mesons can exists at a temperature of 330 MeV and with a  density
of  4  hadrons/$fm^3$.  The  hadrons  will overlap completely and
there identity will be lost. However, we still consider it as our
intention  was  to  compare  frozen   motion   model   with   the
calculations of Srivastava et al \cite{sr96} where such a gas was
considered.  The  situation is better for the viscous hadron gas.
Hadron density is considerably  lower,  it  varies  from  2.7-0.4
hadrons/$fm^3$ their existence can be argued.

\section{Dilepton emission rate}

The  dilepton  emission  rate from the hot hadronic gas have been
computed  by  Srivastava  et  el\cite{sr94b}.  All  the  possible
reaction  channels including the $A_1$ resonance were considered.
The details can be found in ref.  \cite{sr94b,ga94}.  For  lepton
pairs    of    invariant    mass    (M)   and   transverse   mass
$M_T=\sqrt{M^2+p^2_T}$ where $p_T$ is  the  transverse  momentum,
the production rate can be written as,

\begin{equation}
\frac{dN}{dx^4dM^2 dM_T}=\frac{\sigma_{eff}(M)}{2(2\pi)^4}M^2 M_T
K_0(M/T)        \label{6}
\end{equation}

with
\begin{equation}
\sigma_{eff}(M)=\frac{4\pi}{3}\frac{\alpha^2}{M^2}
[1+\frac{2m_l^2}{M^2}]
[1-\frac{4m_l^2}{M^2}]^{1/2}[1-\frac{4m_\pi^2}{M^2}]F_{eff}
\end{equation}

\noindent   where   $F_{eff}$   is   the  effective  form  factor
\cite{comm}. The corresponding result for the emission  from  the
QGP is obtained from the above by replacing $F_{eff}$ by 24/3 and
$m_\pi$ by $m_q$ \cite{sr96}. In writing eq.\ref{6}, we have not
distinguished   between   the   particle   and  fluid  rapidities
\cite{sh94}.

Invariant  mass  distribution  of  lepton  pairs  is  obtained by
convoluting the above equation over the space-time history of the
system. In the HELIOS-3 experiment,  pseudorapidity  rather  than
rapidity  was  measured.  The above rapidity distribution is then
converted into the pseudorapidity distribution  using  the  usual
procedure.   For   comparison  with  experiment  we  require  the
acceptance of dileptons in certain rapidity  window  $[y_1,y_2]$.
For   dileptons  with  invariant  mass  M  and  rapidity  y,  the
acceptance is calculated as \cite{sh94},

\begin{equation}
A(y,M)=\int^{min[1,tanh(y_2-y)]}_{max[0,tanh(y-y_1)]}       f(cos
\theta) d cos \theta
\end{equation}

\noindent  where  $f(x)=1+x^2$  is  the process dependent angular
distribution and is normalised within x=0,1.

The   invariant   mass   distribution  of  lepton  pairs  in  the
pseudorapidity interval $[\eta_1,\eta_2]$ is then obtained as,

\begin{eqnarray}
\frac{dN}{dM}=&&\pi R^2 \frac{\sigma_{eff}(M)}{(2\pi)^4}M^3
\int^{\eta_2}_{\eta_1} d\eta \int^{\tau_f(\eta)}_{\tau_i}
\tau d\tau \nonumber\\
&&\times
\int^{\infty}_{P_{T_{min}}} dP_T A(y,M)
\sqrt{1-\frac{M^2}{M^2_T cosh(y)^2}} P_T K_0(M_T/T)
\end{eqnarray}

In  fig.4,  we have shown the result of our calculation in the PT
scenario (the solid line). It is compared with the calculation of
Srivastava et al \cite{sr96} which  was  obtained  following  the
treatment   of  Sinyukov  et  al  \cite{si91}.  Dilepton  numbers
predicted in the frozen motion model is nearly  a  factor  of  10
lower  than  that  obtained  in  the  treatment of Sinyukov et al
\cite{si91}. The difference between the  two  model  calculations
can be understood. In the treatment of Sinyukov et al \cite{si91}
the  fluid and the particle rapidities are distinguished. For the
HELIOS-3 data Srivastava et  al  \cite{sr96}  limited  the  fluid
rapidity  between  [-1.7,1.7]  and used the Bjorken hydrodynamics
with  dN/dy=225.  They  obtained  the  initial   temperature   as
$T_i=204$  MeV  which  remain fixed throughout the fluid rapidity
region. In the frozen motion model, the  initial  temperature  of
the  fluid  being  is a function of the rapidity and was never so
large  in  the   rapidity   range   covered   by   the   HELIOS-3
collaboration.  Indeed  as  described  earlier,  in some rapidity
ranges, the fluid is in mixed phase,  rather  than  in  pure  QGP
phase. Thus on the average, the initial fluid temperature is much
less  in  the present model than in the model of Srivastava et al
\cite{sr96}. Consequently, production of dileptons is less in the
present model than in their model.

In  fig.5,  the  same results in the NPT scenario is shown. As in
the PT scenario, here  again,  we  find  that  the  treatment  of
Sinyukov  et el\cite{si91} results in more dileptons (by a factor
of 10) than in the frozen motion model. As before, the difference
can be attributed to the larger initial temperature of the  fluid
in  Sinyukov treatment than in the frozen motion model. Thus both
in the PT and the NPT  scenario,  frozen  motion  model  predicts
lesser dileptons than in the model of Sinyukov et al \cite{si91}.

In  fig.  6,  we have compared the invariant mass distribution of
lepton  pairs  as  obtained   by   the   HELIOS-3   collaboration
\cite{ma95}  with  the  dilepton  yields  obtained  in the frozen
motion model. The solid line corresponds to the phase  transition
scenario,  when  the  QGP  is  formed  in  the initial state. The
long-dashed line corresponds to the no phase transition  scenario
with the ideal hadron gas formation as the initial state. We have
also  shown  the  dilepton  yield  obtained  in the scenario when
viscous hadronic gas is formed in the initial state  (the  dashed
line).

All  the  three scenarios underpredict the low mass (M$<$0.8 GeV)
dilepton pairs. As mentioned earlier local fluid approximation is
a good approximation only for high mass dilepton pairs.  For  low
mass  dilepton  pairs, the model predictions may not be reliable.
The low  mass  dileptons  are  essentially  from  the  freeze-out
surface  and  it  is  possible  to  improve  the  fit of low mass
dileptons by increasing the freeze-out temperature.  However,  we
desist  from  such  parameter  fitting.  As  now  well known, the
discrepancy may be more due to neglect of modification of  $\rho$
mass  and  width  in  hot  and dense hadronic medium \cite{li95}.
However, since the present model is not  reliable  for  low  mass
dileptons, we will not discuss this issue further.

In   the   invariant   mass  range  of  1-2  GeV,  the  data  are
underpredicted in all the three scenarios. Thus in all the  three
scenarios the frozen motion model do not produce requisite number
of  lepton  pairs in the invariant mass range 1-2 GeV, indicating
excess dileptons in the experiment.  To  have  quantitative  idea
about the excess dileptons, we define a ratio r as,

\begin{equation}
r=\frac{\text{Integral contents of data histogram}}
{\text{Integral content  of
theory}}
\end{equation}

The  ratio  r  was calculated in the invariant mass range 1-2 GeV
and found to be 10.0, 2.2 and 4.2  for  the  initial  QGP,  ideal
hadronic  gas and the viscous hadronic gas respectively. Dilepton
excess is minimum in the NPT scenario with the ideal  hadron  gas
formation and maximum in the PT scenario.

The  PT  scenario  also  underpredict  the large invariant mass
(M$>$2GeV) data. They are underpredicted by a factor of of 40  or
so. On the otherhand the NPT scenario with the ideal hadronic gas
overpredict the large invariant mass data by a factor of 5. Very
good  description  of  the large mass data is obtained in the NPT
scenario with  the  viscous  hadronic  gas.  As  the  large  mass
dileptons  are essentially from the initial state of the produced
fireball, it seems that it is best described by the  hot  viscous
hadronic gas.

Present analysis indicate that in the frozen motion model, the PT
scenario  do  not  describe  HELIOS-3 data. The NPT scenario also
donot fits the data however overall description is better than in
the PT scenario. This is exactly opposite to the results obtained
by Srivastava  et  al  \cite{sr96}  following  the  treatment  of
Sinyukov  et  al\cite{si91}.  Of the two NPT scenarios considered
e.g. ideal and viscous hadronic gas, the ideal hadronic gas  give
better  description  to the data compared to the viscous hadronic
gas, though the high mass dileptons are overpredicted by a factor
of 5 or so. However, as discussed earlier, in the ideal  hadronic
gas  scenario, the initial temperature of the fluid is very large
($>$300 MeV) and it is difficult to conceive existence of such  a
gas.  Thus  though the hadronic gas gives a better description to
the data, it can not considered seriously. The  viscous  hadronic
gas  on  the  otherhand require much less initial temperature and
can be believed to be in existence.

One  of  the  limitation  of  the present model is the neglect of
baryons. While the assumption of  baryon  free  fireball  may  be
quite  accurate  in  the  central rapidity region, in the forward
rapidity region it is not so. As the HELIOS-3 data  lies  in  the
forward  rapidity  region,  it  is  essential to include baryons.
Inclusion of baryons will introduce  two  complimentary  effects.
The  initial  temperature  of  the fireball will be reduced, as a
result  of  which  dilepton  yield  will  be  decreased.  On  the
otherhand  dilepton  yield  will  be  increased  as the number of
reaction channels are increased. It is possible  that  these  two
complimentary  effects cancels each other and the present results
is not changed significantly \cite{sr96}.

\section{conclusion}

To  summarise, we have analysed the HELIOS-3 dilepton data in the
frozen  motion  model.  Earlier  Srivastava  et  al   \cite{sr96}
analysed  this  data  following  the  treatment of Sinyukov et al
\cite{si91}. They found that formation  of  QGP  as  the  initial
state fits the data. If on the otherhand, hadronic gas is formed,
the  data  could  not  be  fitted. In the frozen motion model, we
obtain entirely different results. Both the PT and  NPT  scenario
could  not fit the data satisfactorily. There is excess dileptons
in the intermediate mass range. Interestingly, we also find  that
the  NPT  scenario  gives  the  better  description  to  the data
compared to the PT scenario. Thus  if  one  can  believe  in  the
existence  of  hot hadronic gas at a temperature in excess of 300
MeV, reasonable description to the data can be obtained. However,
it is difficult to believe that hadrons retain their identity  at
such  large  temperature. The other NPT scenario with the viscous
hadronic gas, do not require very large initial  temperature  and
gives very good description of large mass dileptons. It can be an
alternative   to   the  ideal  hadronic  gas  that  is  generally
considered in literature.

\begin{figure}
\caption{  The  rapidity  distribution of the charged particle in
200 A GeV S+Au collisions as obtained by the WA80  collaboration.
The solid line is a Gaussian fit to the data.}
\end{figure}
\begin{figure}
\caption{Initial  temepratures  ($T_i$)  of the QGP, ideal hadron
gas and the viscous hadron gas as a function of rapidity.}
\end{figure}
\begin{figure}
\caption{Initial  density ($\rho_i$) of the ideal and the viscous
hadronic gas as a function of rapidity.}
\end{figure}
\begin{figure}
\caption{The  invariant  mass distribution of dilepton pairs in
the Phase transition scenario. The solid line corresponds to  the
frozen  motion  model.  The  dotted  lines  are results from ref.
\protect\cite{sr96}.   Cuts   corresponding   to   the   HELIOS-3
experiments are incorporated.}
\end{figure}
\begin{figure}
\caption{The  invariant  mass distribution of dilepton pairs in
the no Phase transition scenario. The solid line  corresponds  to
the  frozen  motion model. The dotted lines are results from ref.
\protect\cite{sr96}.   Cuts   corresponding   to   the   HELIOS-3
experiments are incorporated.}
\end{figure}
\begin{figure}
\caption{The  invariant  mass distribution of dilepton pairs as
obtained by the HELIOS-3 collaboration. The lines corresponds  to
initial QGP, ideal hadron gas and viscous hadron gas.}
\end{figure}

\begin{references}
\bibitem{comm}        We        have        subtracted        the
$\pi^+\pi^-\rightarrow\rho^0\rightarrow e^+e^-$ from the  $F_{eff}$
as  calculated  in ref\cite{sr94b}. This being a dominant reaction
counted twice in ref\cite{sr94b}. Thus the present $F_{eff}$ do not
have the problem of double counting.
\end{references}
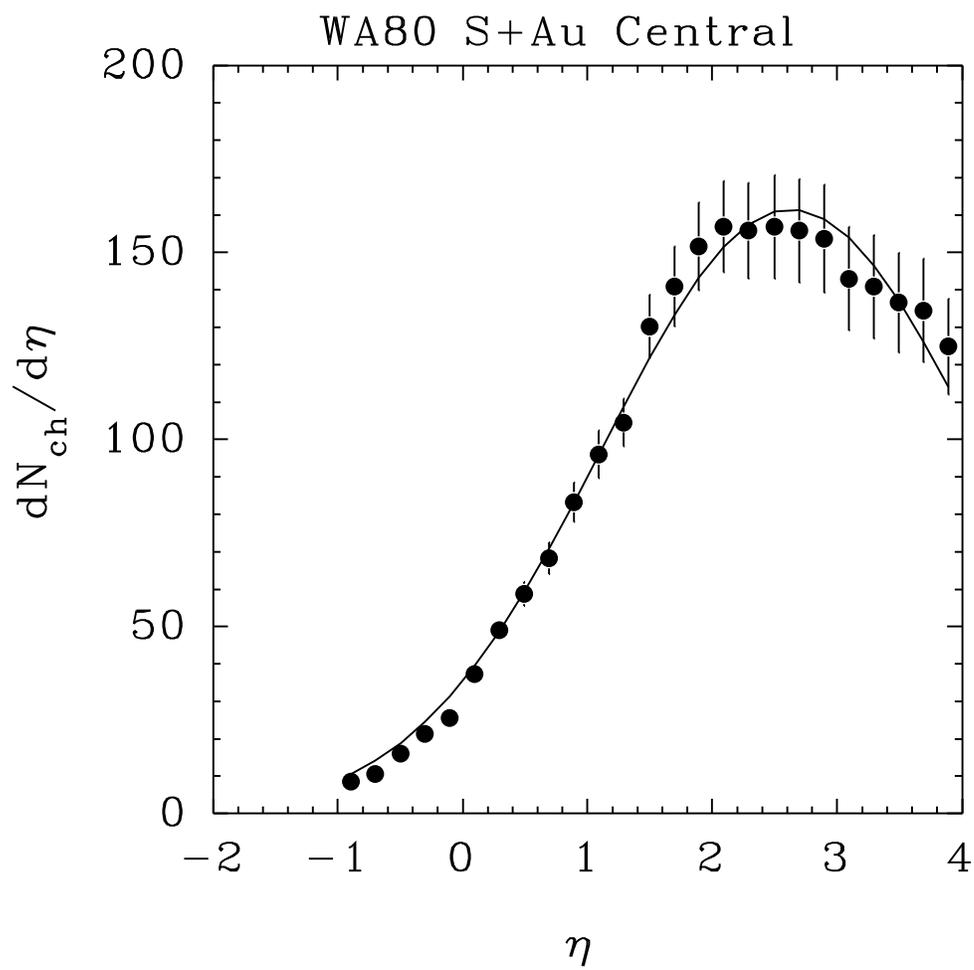
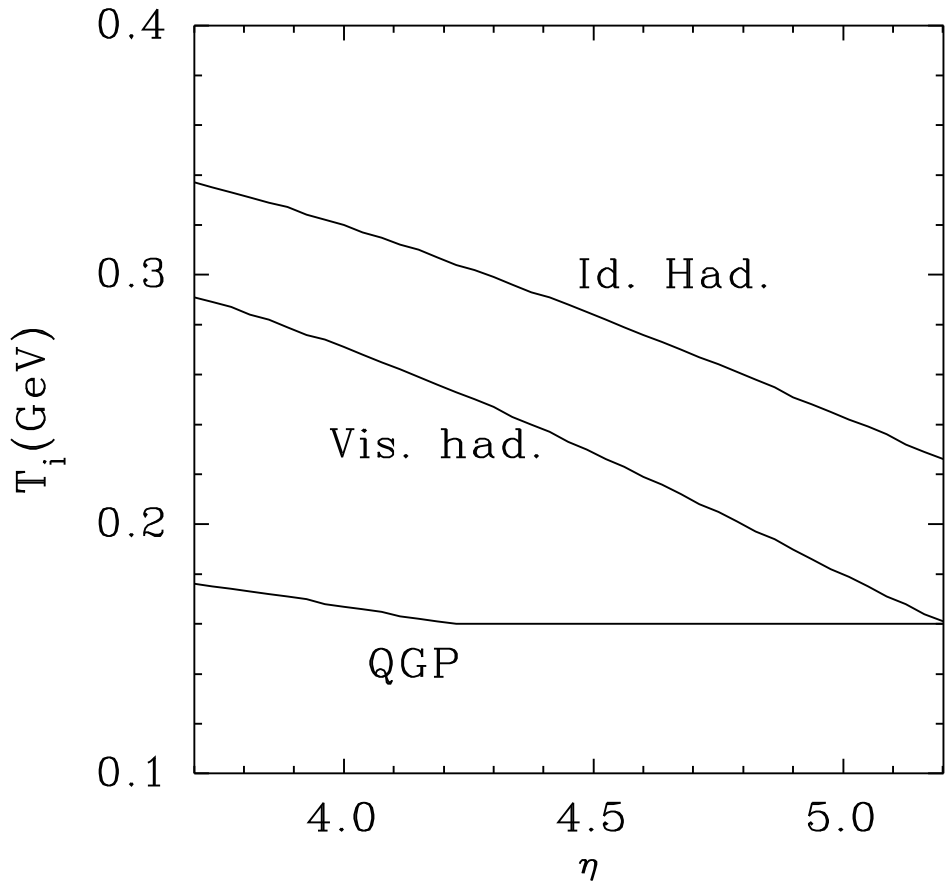
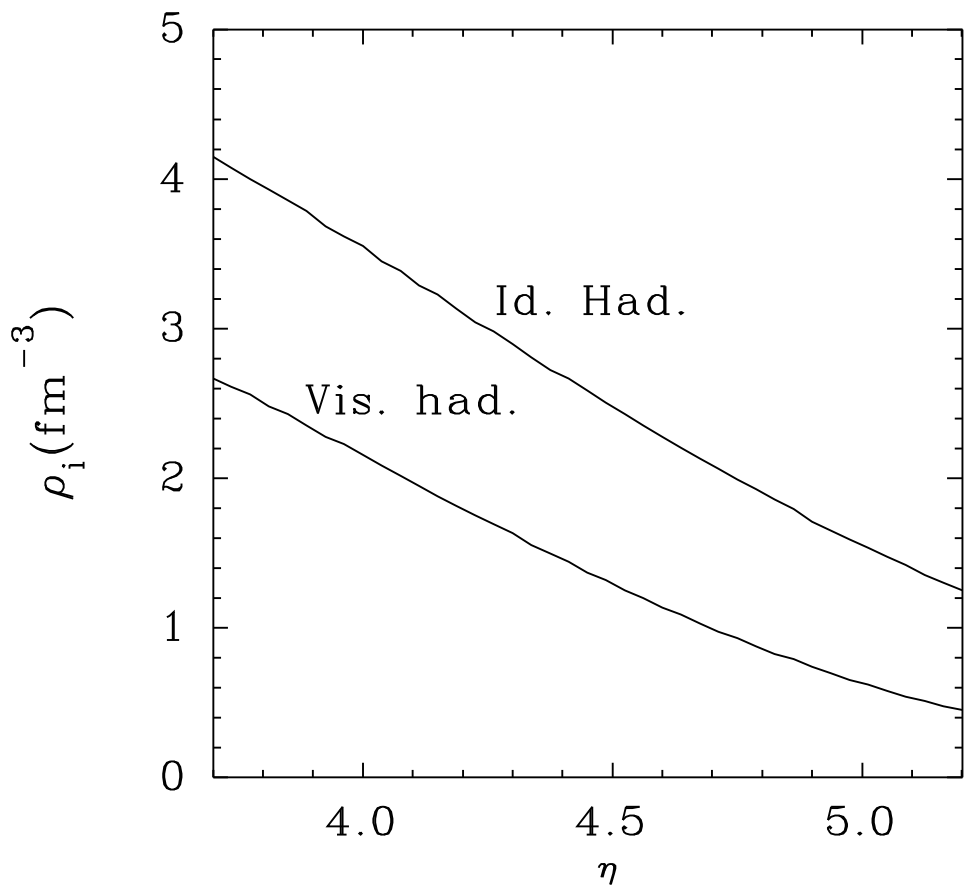
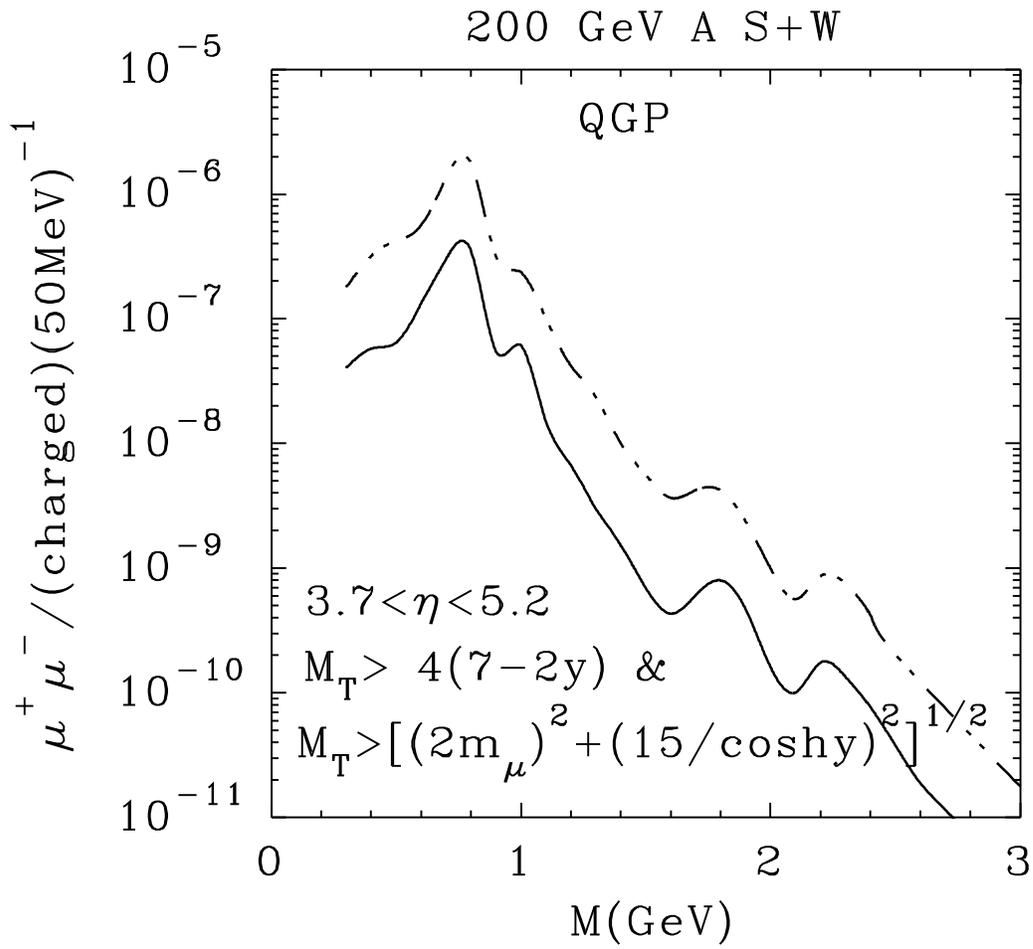
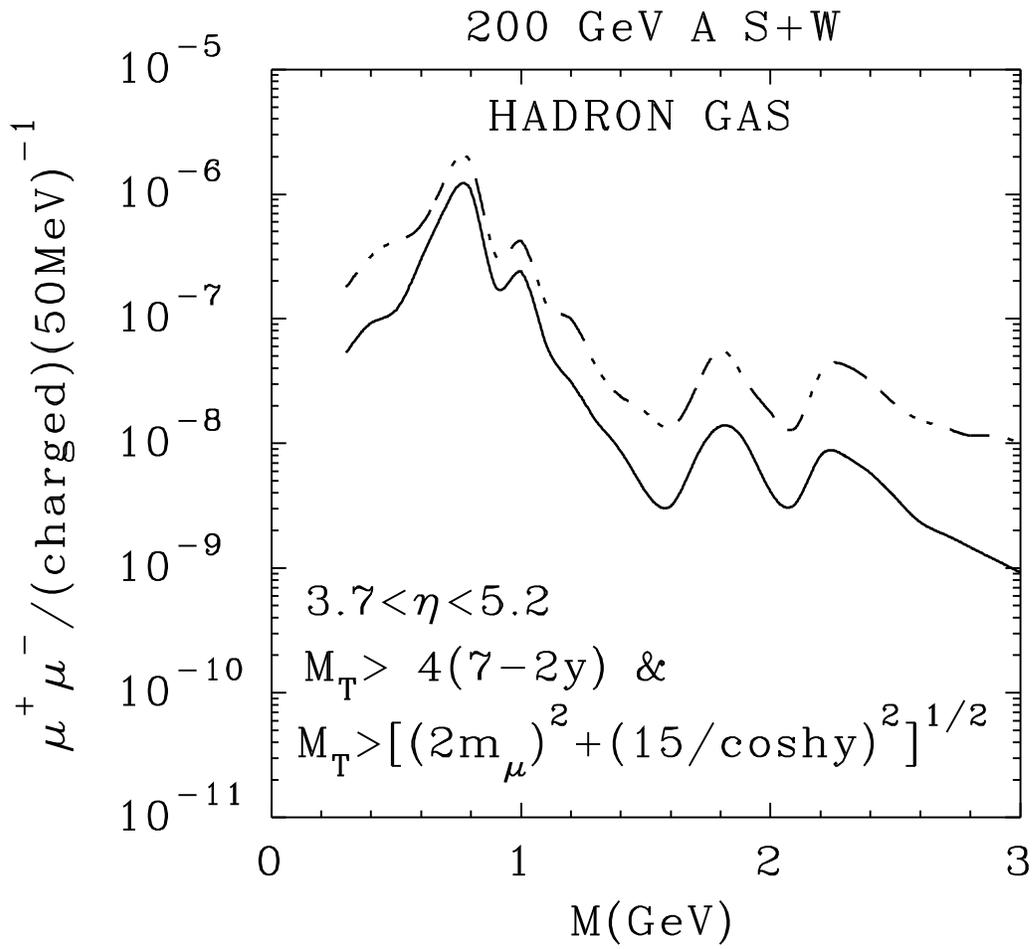
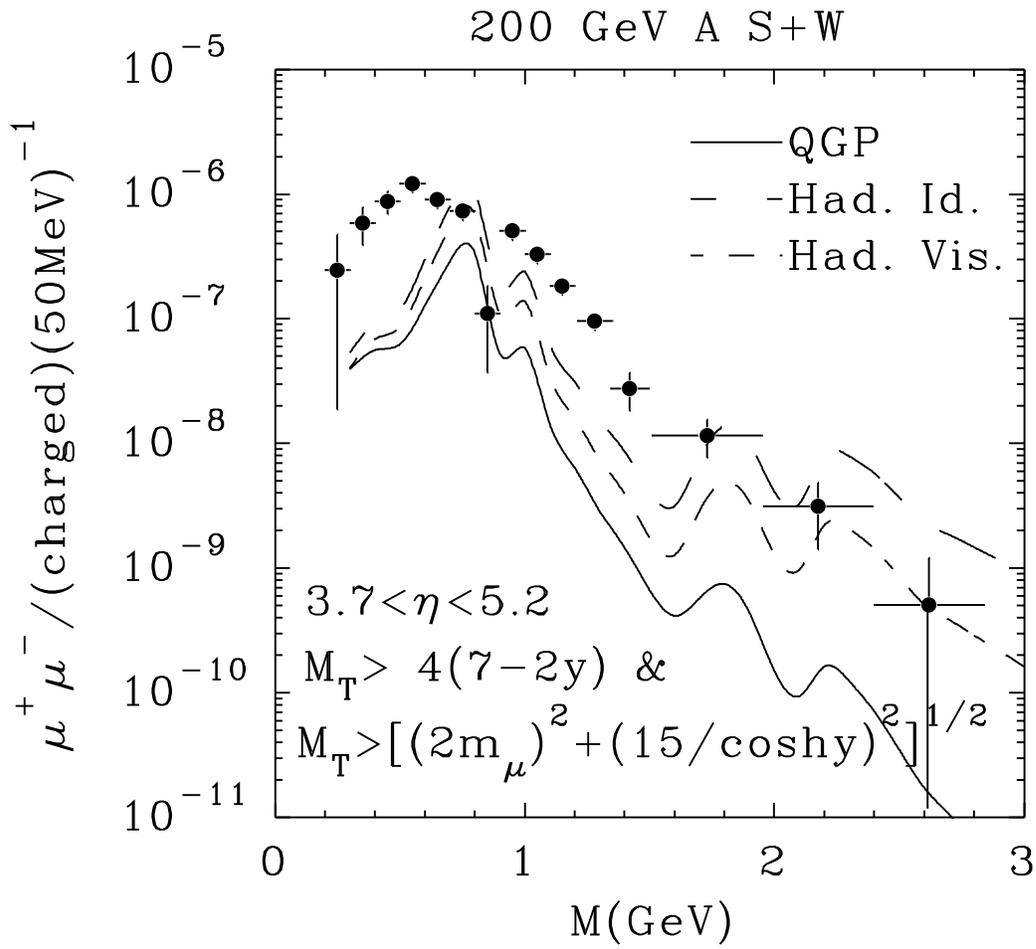
\end{document}